\documentclass[%
 reprint,nofootinbib,
 amsmath,amssymb,
 aps,twocolumn,pra
]{revtex4-2}

\usepackage{graphicx}% Include figure files
\usepackage{hyperref}% add hypertext capabilities
\usepackage{multirow}% For tables
\usepackage{color}
\usepackage{amsfonts}
\newcommand{\eqn}[1]{\begin{eqnarray} #1 \end{eqnarray}}
\newcommand{\tit}[1]{\textit{#1}}
\newcommand{\tbf}[1]{\textbf{#1}}
\newcommand{\trm}[1]{\textrm{#1}}

\newcommand{\zum}[2]{\displaystyle\sum_{#1}^{#2}}

\newcommand{\ket}[1]{| #1 \rangle}
\newcommand{\ketbra}[2]{| #1 \rangle \langle #2 |}

\definecolor{airforceblue}{rgb}{0.36, 0.54, 0.66}

\begin{document}

\title{A quintet of quandaries: five no-go theorems for Relational Quantum Mechanics}

\author{Jacques Pienaar}
\affiliation{
QBism Group, University of Massachusetts Boston, USA.
}

\date{\today}

% \begin{description}
% \item[PACS numbers]

% \pacs{Valid PACS appear here}
% \end{description}

\begin{abstract}
Relational quantum mechanics (RQM) proposes an ontology of relations between physical systems, where any system can serve as an `observer' and any physical interaction between systems counts as a `measurement'. Quantities take unique values spontaneously in these interactions, and the occurrence of such `quantum events' is strictly relative to the observing system, making them `relative facts'. The quantum state represents the objective information that one system has about another by virtue of correlations between their physical variables. The ontology of RQM thereby strives to uphold the universality and completeness of quantum theory, while at the same time maintaining that the actualization of each unique quantum event is a fundamental physical event. Can RQM sustain this precarious balancing act? Here we present five no-go theorems that imply it cannot; something has to give way.
\end{abstract}

\maketitle

\section{Introduction}

The relational interpretation of quantum mechanics (RQM) was first proposed by Carlo Rovelli in 1996 \cite{ROVELLI_96} and has since been further developed by Rovelli and other authors \cite{SEP_RQM,RQM_STABLE_FACTS,RQM_BROWN,RQM_BUNDLE,RQM_CALOSI,RQM_DORATO,RQM_EPR,RQM_KRISMER,RQM_LAUDISA_EPR,RQM_LAUDISA_OPEN,RQM_BITBOL,RQM_LOCALITY,RQM_RUYANT,RQM_TRASSINELLI,RQM_VANF,RQM_YANG,RQM_WOOD,RQM_ROVELLI_16,RQM_ROVELLI_BIRDS}. One of the most striking features of RQM is its adherence to the principle of \tit{relative facts}: quantum events correspond to single, unique elements of reality which are \tit{relative to} the observer who measures them. The existence or not of a quantum event cannot be established absolutely: the fact of a measurement having produced an outcome is fundamentally relational. In particular there are situations -- most notably the Wigner's friend thought experiment -- in which a quantum event which is a fact for one observer (the friend) cannot be taken to be a fact for another observer (Wigner). 

RQM has the distinction of being the oldest interpretation which has consistently endorsed relative facts\footnote{Many worlds interpretations, while possibly in some sense `relational', do not endorse the uniqueness of relative facts.}. The second oldest is QBism \cite{QB_FS,QBISM_HERO,BCS_EnRoute,FAQBism,FuchsNWB,Fuchs_PR,Fuchs_PERIMETER,QB_FDR2014} which was explicitly articulated only in 2010 \cite{Fuchs_PERIMETER}; and Brukner has endorsed relative facts since at least 2015 \cite{BW_BRUKNER,BRUKNER_17,BW_BRUKNER_NATURE}.

Recently, a series of papers by various authors have combined insights from Bell's and Wigner's famous thought experiments to create new quantum no-go theorems\footnote{Matthew Leifer calls them `Bell-Wigner mashups' \cite{LEIFER_WIGNER}.} for interpretations of quantum theory \cite{BW_BAUMANN_WOLF,BW_BONG,BW_BRUKNER,BW_COMMENT,BW_FR,BW_HEALEY,BW_PROIETTI,BW_CAVALCANTI,QB_FELLOW,BW_BRUK_BAUM,BW_BRUKNER_NATURE}. In all these works, the endorsement of relative facts appears to provide a potentially attractive escape route from the no-go theorems. Therefore the time is ripe to look more closely at interpretations which claim this principle as a central tenet, particularly RQM and QBism.

While RQM and QBism agree on a number of points, the incompatibilities between the two has not been much explored in the literature. Therefore a companion paper, Ref.\ \cite{QBISM_COMPARISON}, will be devoted to pointing out key differences between RQM and QBism, from a QBist standpoint. In contrast, the present work is not a comparison but a critique: we seek to expose essential weaknesses of RQM, presenting them in the form of five `no-go' results.

To avoid attacking a straw man, we will accept the core principles of RQM at face value and aim to show that there are internal conflicts among them. As such, we do not adopt a QBist viewpoint here. Instead, we adopt a more inclusive attitude that is broadly aligned with what Matt Leifer has called `Copenhagenish' interpretations \cite{LEIFER_WIGNER}, including RQM. In particular, we accept the idea that measurements are associated to quantum events that are single, unique and fundamental to the ontology; that wavefunctions have a broadly epistemic character; and that the quantum formalism is both complete and universal (for more details, see Ref.\ \cite{LEIFER_WIGNER} and our companion paper Ref.\ \cite{QBISM_COMPARISON} ). 

Our approach differs from that of another recent criticism of RQM (Ref.\ \cite{RQM_MOS}). The authors of that work follow Bell's famous essay ``Against `measurement' " \cite{BELL_AM} in declaring that measurements ``should not appear as primitives in a theory that aims at being fundamental", because this would lead to ``an unacceptable vagueness at its core" \cite{RQM_MOS}. In response, Rovelli argued that the critique was ``an exercise in misunderstanding" because the authors were ``interpreting RQM on the basis of assumptions rejected by RQM" \cite{RQM_VS_MOS}. We note that since all Copenhagenish interpretations treat measurement as a fundamental notion, the general standpoint of Ref. \cite{RQM_MOS} is antagonistic to all of them. This motivates us to adopt a standpoint that is sympathetic to Copenhagenish interpretations generally, in order to better identify those afflictions more specific to RQM.

Our task is complicated by the fact that not all contributors to the RQM literature have shared the same understanding of it, and even Rovelli has acknowledged that his own understanding of RQM has evolved over time \cite{RQM_VS_MOS}. We shall therefore base our arguments on the version of RQM as it is currently endorsed by Rovelli, taking our primary literature to be the published and unpublished writings of Rovelli (possibly with co-authors).

Finally, we shall not address the issue of whether RQM admits a ``local" ontology here. This is partly because the issue has been amply covered elsewhere \cite{RQM_LOCALITY, RQM_PIENAAR_COMMENT, RQM_RUYANT, RQM_MOS}, but also because \tit{any} interpretation that endorses relative facts is likely to face similar difficulties in defining ``locality" and therefore the problem is not entirely unique to RQM\footnote{For instance, a pertinent criticism of QBism's claim to locality can be found in Sec.\ 3.2. of Ref.\ \cite{BW_CAVALCANTI}.}.

The paper is organized as follows. In Sec.\ \ref{sec:principles} we review certain key principles of RQM that will be essential to our subsequent arguments. We divide our arguments into two main categories: Sec.\ \ref{sec:not_relational} presents arguments challenging the supposed analogy between relations in RQM and relations in relativistic physics, while Sec.\ \ref{sec:not_objective} presents arguments targeting the supposed `objectivity' of relations in RQM. Each section contains separate subsections that present the main arguments as no-go theorems.

\section{Key claims of RQM \label{sec:principles}}

In this section we review selected key features of RQM that will be essential to our arguments, which we present as a set of six `claims' labeled \tbf{RQM:1}--\tbf{6}. Some of these claims are simply re-statements of core principles of RQM, while others are indirect consequences of the core principles. All claims are supported by the primary literature and so we expect them to be mostly uncontroversial to proponents of RQM.

The core thesis of RQM consists of the following main elements. A ``measurement" is nothing more than a \tit{physical interaction} between two systems, in which either system can stand in the role of \tit{observer} while the other takes the role of \tit{observed system}, and in which the value of some physical variable of the observed system spontaneously acquires a unique value \tit{relative to} the designated observer. The spontaneous actualization of a physical variable relative to an observer is called a \tit{relative fact} or \tit{quantum event}\footnote{Rovelli tends to eschew the terms ``measurement" and ``measurement outcome" in order to avoid the unwanted interpretational baggage that often accompanies these terms \cite{RQM_VS_MOS}.} (we use these terms interchangeably). Quantum events are strictly physical occurrences that require nothing of the observer beyond that it be a physical system. Furthermore, the occurrence of a quantum event is said to establish a \tit{relation} between the observer and observed system, in which the former is said to have `acquired information' about the latter. Finally, the question of whether or not a quantum event has absolutely occurred independently of any system is not a well-posed question in RQM: it can only be answered relative to some system chosen to play the role of observer. We call this the principle of \tit{relative facts}.

Within this broad picture, we now identify some more specific claims. Since ``measurement" or ``observation" is a strictly physical process, an observer is necessarily a physical system. It is a core principle of RQM that the converse is also true, namely:

\begin{quote}
    \tbf{RQM:1. Any system can be an observer.} Any physical system can play the role of an observer in a physical interaction.
\end{quote}

An important concept in RQM is the \tit{perspective} of an observer, defined as the ``ensemble of all events relative to [the observer], together with the probabilistic predictions these entail" \cite{SEP_RQM}. The latter probabilistic predictions are encoded in the quantum state:

\begin{quote}
[The quantum state is] a mathematical device that refers to two systems, not a single one. It codes the values of the variables of the first that have been actualised in interacting with the second [\dots] [It] therefore codes anything we can predict regarding the future values of these variables, relative to the second system. The state $\Psi$, in other words, can be interpreted as nothing more than a compendium of information assumed, known, or gathered through measurements, determined entirely by a specific history of interactions: the interactions between the system and a second ‘observing’ system. --- \cite{SEP_RQM}.
\end{quote}

A core principle of RQM is that quantum theory is ``complete", in the sense that there is no ``deeper underlying theory that describes what happens `in reality' " and hence there are no \tit{hidden variables} in RQM's ontology beyond ``the relevant information that systems have about each other" \cite{ROVELLI_96} (p.1650). Intuitively, the quantum state of a system relative to an observer is determined only by the quantum events in the observer's perspective\footnote{Note that the quantum events relevant for determining a system's state are not only interactions with the system itself, but also interactions with other systems that have previously interacted with it.}. To say that it is \tit{complete} means that there is no other information which is in principle accessible to the observer which would permit a more accurate prediction about the observer's subsequent interactions with the system. We may state this as the following claim:

\begin{quote}
    \tbf{RQM:2. No hidden variables.} Any variable that exists in the observer's causal past and which is relevant to predictions about future quantum events relative to the observer must be a quantum event contained in their perspective.\\ 
\end{quote}

The absence of hidden variables means that the only elements of reality in RQM are the relations between systems (as manifested in quantum events), and possibly the systems themselves\footnote{For more about whether systems are real, see Rovelli's remarks at the end of Sec.\ X in \cite{RQM_VS_MOS}.}. Rovelli characterizes this picture as ``a sparse ontology of (relational) quantum events happening at interactions between physical systems"\cite{RQM_ROVELLI_BIRDS}. 

In RQM it is stipulated that relations are \tit{intrinsic} to pairs of systems, meaning that no more than two are required in order to define a relation. We can make this more precise by re-formulating \tbf{RQM:2} in a more symmetric way that does not depend on which system is playing the role of the observer:

\begin{quote}
    \tbf{RQM:3. Relations are intrinsic.} The relation between any two systems $\mathcal{A}$ and $\mathcal{B}$ is independent of anything that happens outside these systems' perspectives. In particular, the state of $\mathcal{B}$ relative to $\mathcal{A}$ depends only upon $\mathcal{A}$'s observation of $\mathcal{B}$ and $\mathcal{A}$'s past history of interactions (similarly for the state of $\mathcal{A}$ relative to $\mathcal{B}$). 
\end{quote}

Note that the probabilities that comprise the quantum state are necessarily relational, i.e.\ it is meaningless to talk about probabilities for quantum events except as probabilities relative to some observer. This implies that even the question of whether two systems have interacted (or not) is itself a fact relative to an observer. As Rovelli puts it (our emphasis):

\begin{quote}
[T]he fact that a variable in a system $\mathcal{O}$ has information about a variable in a system $\mathcal{S}$ means that the variables of $\mathcal{S}$ and $\mathcal{O}$ are correlated, meaning that a third observer $\mathcal{P}$ has information about the coupled $\mathcal{S}$-$\mathcal{O}$ system that allows her to predict correlated outcomes between questions to $\mathcal{S}$ and questions to $\mathcal{O}$. \tit{Thus correlation has no absolute meaning}, because states have no absolute meaning, and must be interpreted as the content of the information that a third system has about the $\mathcal{S}$-$\mathcal{O}$ couple. --- \cite{ROVELLI_96}  (p.1669).
\end{quote}

Significantly, this extends also to comparisons of measurement results between systems. That is to say, in RQM it is only meaningful to compare the results of measurements between two observers by invoking a third observer relative to whom the comparison is described as a physical interaction. Smerlak and Rovelli are quite explicit (emphasis in original):

\begin{quote}
    The founding postulate of RQM stipulates that we shall not deal with properties of systems in the abstract, but only of properties of systems relative to \tit{one} system. In particular, we can never juxtapose properties relative to different systems. [\dots] In other words, RQM is \tit{not} the claim that reality is described by the \tit{collection} of all properties relative to all systems. This collection is assumed not to make sense. Rather, reality admits one description per (observing) system, each being internally consistent. [\dots] Contradiction emerges only if, against the main stipulation of RQM, we insist on believing that there is an absolute, external account of the state of affairs in the world, obtained by juxtaposing actualities relative to different observers. --- \cite{RQM_EPR} (p.441).
\end{quote}

We summarize this as the following claim:

\begin{quote}
    \tbf{RQM:4. Comparisons are relative to one observer.} It is meaningless to compare the accounts of any two observers except by invoking a third observer relative to which the comparison is made.
\end{quote}

A measurement in RQM is considered nothing more than the acquisition of mutual information (in the sense of Shannon's definition) between the \tit{physical variables} of two systems. It is therefore important to know what is meant by a \tit{physical variable}, and whether this includes all Hermitian operators or not. Rovelli clarifies (our emphasis):

\begin{quote}
    [G]iven an arbitrary state of the coupled $\mathcal{S}$-$\mathcal{O}$ system, there will always be a basis in each of the two Hilbert spaces which gives the bi-orthogonal decomposition, and therefore which defines an [operator on the Hilbert space of $\mathcal{S}$-$\mathcal{O}$] for which the coupled system is an eigenstate. But this is of null practical nor theoretical significance. \tit{We are interested in certain self-adjoint operators only}, representing observables that we know how to measure; for this same reason, we are only interested in correlations between certain quantities: the ones we know how to measure. --- \cite{ROVELLI_97}, (p.8).
\end{quote}

As pointed out in Ref.\cite{RQM_MOS}, Rovelli's appeal to quantities that `we know how to measure' fails to meet RQM's own mandate to define these quantities in strictly physical terms. Even assuming this problem could be somehow overcome, any strictly physical definition of the physical variables must at least include those which are familiar to us as measurable quantities, the values of which must then correspond to the eigenvalues of a Hermitian operator on the system's Hilbert space. This is enough to establish \tit{necessary} conditions for a variable to be physical. 

In fact, we argue that in RQM these conditions must also be \tit{sufficient}. For according to Rovelli, there is nothing else that could single out one physical variable as having a more fundamental status than any other. Rovelli writes:

\begin{quote}
I shall use here a notion of information that does not require distinction between human and nonhuman observers, systems that understand meaning or do not, very complicated or simple systems, and so on. [\dots] Information expresses the fact that a system is in a certain configuration, which is correlated to the configuration of another system (information source). [\dots] [In particular] we do not distinguish between ``obtained" correlation and ``accidental" correlation (if the pointer of the apparatus indicates the correct value of the spin, we say that the pointer has information about the spin, whether or not this is the outcome of a ``well-thought" interaction) [\dots]. --- \cite{ROVELLI_96}, (p.1653-54).  
\end{quote}

This implies that, when deciding whether a correlation between variables of two systems constitutes a measurement, it is sufficient that the variables should be \tit{physical variables}. Any further considerations, say, of meaningfulness to human scientists, complexity, etc, are irrelevant. Consequently:

\begin{quote}
    \tbf{RQM:5. Any physical correlation is a measurement.} Suppose an observer measures a pair of systems and thereby assigns them a joint state which exhibits perfect correlations between some physical variables. Then the two systems have measured each other (entered into a measurement interaction) relative to the observer, and the physical variables play the roles of the `pointer variable' and `measured variable' of the systems\footnote{Which variable gets assigned to which role depends on which of the two systems we take as the observer.}.
\end{quote}

This leads us at last to the `Wigner's friend' thought experiment: a system $\mathcal{F}$ (the friend) measures another system $\mathcal{S}$, as described by a third system $\mathcal{W}$ (Wigner) who does not participate in the measurement, but is still able to verify that it occurred, and thereby assigns an entangled state to $\mathcal{F}$-$\mathcal{S}$. By \tbf{RQM:5} this state exhibits perfect correlations in some physical variables, say $F_X$ and $X$ of $\mathcal{F}$ and $\mathcal{S}$ respectively.

The primary literature repeatedly emphasizes that although observers can assign different states to the same system, it must always be possible for observers to \tit{agree} by measuring each other in the appropriate `pointer' basis. For example, in the Wigner's friend scenario just described, $\mathcal{W}$ can perform a measurement on $\mathcal{F}$-$\mathcal{S}$ that corresponds to ``checking what outcome $\mathcal{F}$ obtained". It is important to emphasize that the value which $\mathcal{W}$ reads on $\mathcal{F}$'s pointer variable $F_X$ here is supposed to represent a fact previously recorded by $\mathcal{F}$, that is to say, a fact which can be said to have existed \tit{relative to $\mathcal{F}$} prior to $\mathcal{W}$'s measurement.
For instance, Laudisa \& Rovelli write (our notation and emphasis):

\begin{quote}
     [I]magine experimenter $\mathcal{F}$ measures the spin of the electron $\mathcal{S}$, and writes the value of this spin on a piece of paper. In principle, experimenter $\mathcal{W}$ can devise an experiment where she can detect an effect due to interference between the two branches where the spin of the electron (and the text) have one or the other value. But if $\mathcal{W}$ measures the spin and reads the piece of paper, she will find that experimenter $\mathcal{F}$ \tit{has seen} the same spin as herself. --- \cite{SEP_RQM} 
\end{quote}
The ``has seen" makes it clear that the pointer variable -- in this example a piece of paper\footnote{Note that in light of \tbf{RQM:5}, the pointer variable here ought not to have any special status by virtue of being a ``piece of paper", i.e.\ something that you or I would intuitively recognize as a tool for keeping records. Other physical variables might serve equally well as apparatus pointers.} -- represents a record of what $\mathcal{F}$ has measured previously \tit{in her own perspective}. Summarizing,

\begin{quote}
    \tbf{RQM:6. Shared facts.} In the Wigner's friend scenario as outlined above, if $\mathcal{W}$ measures $\mathcal{F}$ to `check the reading' of a pointer variable (i.e.\ by measuring $\mathcal{F}$ in the appropriate `pointer basis'), the value he finds is necessarily equal to the value that $\mathcal{F}$ recorded in her account of her earlier measurement of $\mathcal{S}$.
\end{quote}

This concludes our survey of the relevant features of RQM. In what follows the claims \tbf{RQM:1}-\tbf{6} will be turned against each other in a series of dilemmas and multi-lemmas. 

\section{Challenges to the relativistic analogy of RQM \label{sec:not_relational}}

At the core of every interpretation of quantum mechanics is the desire to de-mystify or dissolve the apparent paradoxes of quantum theory. The common strategy for doing this is to demonstrate that most of quantum theory's counter-intuitive predictions can be explained by certain classical intuitions, plus a residue of one or two essentially quantum postulates. For RQM, this means arguing that quantum theory can be understood by thinking of physical reality as composed of \tit{relational} quantities analogous to, but more general than, classical relativistic relations.

Proponents of RQM often make analogies to classical relativity even while emphasizing that such analogies are imperfect and limited. For as long as the analogy holds even partially, RQM can present quantum theory as a natural continuation and generalization of classical relational intuitions, suitably modified to account for a few empirically inspired postulates about the finiteness and intrinsic randomness of quantum systems \cite{ROVELLI_96}. Seen in this light, quantum theory is not a rejection of classical intuitions, but a dramatic confirmation of classical \tit{relational} intuitions. Without RQM's conceptual analogy to classical relativistic relations, RQM would therefore lose its core motivation as an interpretation.

But is the analogy really justified?

Let us begin by recalling what it means for some property $R$ to be a \tit{relational} quantity in a classical sense. Following our intuitions from relativity theory, $R$ must be a joint property of at least two systems; let us call them $\mathcal{A}$ and $\mathcal{B}$. Intuitively, $R$ can be decomposed as the value of some \tit{non-relational} property of $\mathcal{A}$ as measured relative to the \tit{frame} of $\mathcal{B}$ (or vice-versa). The idea is that this splitting of $R$ into \tit{relata} -- a value relative to a frame -- has no fundamental significance, and it is only the pure relation $R$ which has fundamental meaning. This means that the pure relation $R$ is itself a \tit{frame-independent} or \tit{absolute} quantity, that is to say, we can talk about $R$ without having to introduce any further system to serve as a frame for $R$. The absoluteness of classical relations manifests in three distinctive ways:
\begin{itemize}
\item\tbf{A1.} The statement that a relation $R$ obtains between $\mathcal{A}$ and $\mathcal{B}$ is a frame-independent statement;
\item\tbf{A2.} The quantitative magnitude of $R$ is frame-independent;
\item\tbf{A3.} The set of all relations defined for all pairs of systems can be taken together to form a unified relational structure that can be defined independently of any particular frame.
\end{itemize}

For a concrete example, let $\mathcal{A}$ and $\mathcal{B}$ be systems localized in Minkowski space-time, and let $R$ be the invariant space-time interval between them. The fact that such an interval exists is of course frame-independent, as is its magnitude. And the set of intervals between all systems in space-time form a unified substructure of the space-time metric, which we can contemplate as a geometric object without reference to any particular frame.

The point is that relations in classical relativity are \tit{absolute}. The great strength of classical relativity theory did not come from its decision to boldly embrace the relational aspects of space and time; on the contrary, its strength derives from the identification of \tit{new absolutes} -- notably the invariant intervals in special relativity and the Riemannian manifolds of general relativity -- which provided the bedrock upon which new physics could be built.

It is therefore ironic that the very feature which RQM touts as the key \tit{relational} aspect of quantum theory, namely the principle of relative facts, actually excludes \tit{every single one} of the three properties of classical relations listed above. It is easy to see why: the principle of relative facts implies that the very existence of a relation in RQM is relative to an observer, hence directly excluding A1. Since absoluteness of the magnitude of a relation presupposes its absolute existence, A2 follows suit. And since the absoluteness of the set of existent relations also presupposes the absolute existence of the individual relations, A3 similarly follows suit. \tit{In short, no interpretation that endorses fact relativity can endorse the absoluteness of relations, A1-A3}.

The primary literature on RQM makes no attempt to hide this fact; indeed we have already seen examples of how RQM explicitly rejects the absoluteness of values of quantum events (hence A2) as well as the absoluteness of correlations between quantum events (hence A1). There are also numerous examples in the literature where A3 is rejected, for instance in \cite{ROVELLI_96} Rovelli writes (emphasis in original):

\begin{quote}
There is no way to ``exit" from the observer-observed global system. 
\tit{Any observation requires an observer} [\dots] There is no description of the universe ``\tit{in toto}," only a quantum-interrelated net of partial descriptions. --- \cite{ROVELLI_96} (p.1669-74).
\end{quote}

Whereas classical relativity abandoned absolute spatial distances and temporal durations in favour of new geometric absolutes, RQM seems committed to abandoning absolutes \tit{tout court}. But if that is the case, then what precisely is salvaged of classical intuitions? Paradoxically, despite the explicit rejection of absolutes, the literature on RQM is peppered with statements like the following:

\begin{quote}
The state of a physical system is the net of the relations it entertains with the surrounding systems. The physical structure of the world is identified as this net of relationships. --- \cite{SEP_RQM}
\end{quote}

Are we supposed to understand this ``net of relationships" as being an absolute network? Of course we cannot, if RQM is to be consistent. But in order to make it consistent with RQM's own principles, we would have to write:

\begin{quote}
The physical structure of the \tbf{observer's} world is identified as this net of relationships \tbf{relative to the observer}.
\end{quote}

But then, of course, the statement can no longer be construed as an absolute statement about `the world'. Here is another:

\begin{quote}
I maintain that in quantum mechanics, ``state" as well as 
``value of a variable" -- or ``outcome of a measurement" -- are relational notions in the same sense in which velocity is relational in classical mechanics. We say ``the object $\mathcal{S}$ has velocity $v$" meaning ``with respect to a reference object $\mathcal{O}$." Similarly, I maintain that ``the system is in such a quantum state" 
or ``$q = 1$" are always to be understood ``with respect to the reference $\mathcal{O}$."  In quantum mechanics \tit{all} physical variables are relational, as velocity is. --- \cite{ROVELLI_96}
\end{quote}

We beg to differ: the relative velocity between $\mathcal{S}$ and $\mathcal{O}$ has a frame-independent geometric interpretation; the quantum state of $\mathcal{S}$ relative to $\mathcal{O}$ has no analogous interpretation (because of the rejection of A2).  Aside from the minimum requirement of indexing quantities to frames, \tit{the relations in RQM are nothing at all like classical relativistic relations}. 

In order to drive this point home as thoroughly as possible, we now present two no-go theorems that exhibit strong dis-analogies between `relative facts' on the one hand, and classically relative quantities like velocity on the other. 

\subsection{No-go theorem 1 \label{sec:nogo1}}

In classical relativity, whenever there is a relation $R$ defined between two systems $\mathcal{F}$ and $\mathcal{S}$, no matter what its value might be, it is always clear which physical quantity it refers to, eg.\ we can say that it is either the relative position, or the relative momentum between $\mathcal{F}$ and $\mathcal{S}$. At first glance, RQM seems to have the same feature, namely, whenever a correlation is established between two systems, it is usually stipulated that this correlation exists between certain definite physical variables of the two systems. However, there are situations in RQM in which a relation $R$ is established relative to some observer, such that either $R$ does not correspond to any definite physical quantities, or else $R$ corresponds to simultaneously incompatible measurements. We formulate this as a dilemma:

\begin{quote}
    \tbf{Dilemma:} Suppose a system $\mathcal{F}$ has measured $\mathcal{S}$, and this fact is verified by a third system $\mathcal{W}$ who measures $\mathcal{F}$-$\mathcal{S}$. Then there exist situations in which one of the following must be true:\\
(i) $\mathcal{F}$ has measured $\mathcal{S}$ simultaneously in incompatible bases, relative to $\mathcal{W}$;\\
(ii) The basis in which $\mathcal{F}$ has measured $\mathcal{S}$ is indeterminate relative to $\mathcal{W}$.
\end{quote}

\tit{Proof:} Let $\mathcal{S}$ and $\mathcal{F}$ be physical systems with associated Hilbert spaces $\mathcal{H}_{\mathcal{S}}$ and $\mathcal{H}_{\mathcal{F}}$ respectively, and suppose that $\mathcal{F}$ has measured $\mathcal{S}$. According to RQM, this means that some physical variable $X$ of $\mathcal{S}$ has become correlated with some physical ``pointer variable" $F_X$ of $\mathcal{F}$. 

Moreover, since these are `physical variables', the values $\{x_i \}$ of $X$ (resp. $\{ Fx_{i} \}$ of $F_X$) correspond to eigenvalues of some Hermitian operator on the Hilbert space of $\mathcal{S}$ (resp. $\mathcal{F}$), and we can associate $X$ (resp. $F_X$) to an orthonormal basis $\{ \ket{x_j} : j=1,\dots,D_{\mathcal{S}} \}$ of $X$ (resp. $\{ \ket{Fx_{i}} :i=1,\dots,D_{\mathcal{F}}  \}$ of $F_X$). Hence $\{ \ket{Fx_{i}} \ket{x_j} \}_{\forall i,j }$ is an orthonormal basis for $\mathcal{H}_{\mathcal{F} \mathcal{S}} := \mathcal{H}_{\mathcal{F}} \otimes \mathcal{H}_{\mathcal{S}}$.

Now consider the state of $\mathcal{F}$-$\mathcal{S}$ relative to the third system $\mathcal{W}$. Any such state can be expanded in the $\{ \ket{Fx_{i}}\ket{x_j} \}$ basis:
\eqn{ \label{eqn:basis_gen}
\ket{\Psi}_{\mathcal{F}\mathcal{S}} = \zum{i,j}{}  \, \alpha_{ij} \, \ket{Fx_{i}}_{\mathcal{F}} \ket{x_j}_{\mathcal{S}} \, .
}
Since $X$ and $F_X$ are correlated, we generally expect $\ket{\Psi}_{\mathcal{F}\mathcal{S}}$ to be entangled in the $\{ \ket{Fx_{i}}\ket{x_j} \}$ basis. In the case where $X$ and $F_X$ are only partially correlated, Rovelli interprets this as representing some probability of the measurement being complete or incomplete\footnote{See Ref.\ \cite{ROVELLI_97}, p.8-9: ``there is no half-a-measurement; there is probability one-half that the measurement has been made! [\dots] Imperfect correlation does not imply no measurement performed, but only a smaller than 1 probability that the measurement has been completed".}. For the sake of argument, let us assume the measurement has been completed, in which case there are perfect correlations between $X$ and $F_X$, hence $\alpha_{ij} := \alpha_{i}\delta_{ij}$ and $\ket{\Psi}_{\mathcal{F}\mathcal{S}}$ is an entangled state of Schmidt form:
\eqn{ \label{eqn:basis0}
\ket{\Psi}_{\mathcal{F}\mathcal{S}} = \zum{i}{}  \, \alpha_{i} \, \ket{Fx_{i}}_{\mathcal{F}} \ket{x_i}_{\mathcal{S}} \, .
}
(Note that this decomposition is not unique if the $\alpha_i$ are degenerate; this will be important shortly). Let us assume that there exists another pair of physical variables $Y$ of $\mathcal{S}$ (resp. $F_Y$ of $\mathcal{F}$) which corresponds to an orthonormal basis $\{ \ket{Fy_{m}}\ket{y_n} \}$ of $\mathcal{H}_{\mathcal{F} \mathcal{S}}$. Assume further that the new bases are incompatible with the old ones, i.e.\ they are not simply permutations of one another (we will soon give an explicit example).

Due to the non-uniqueness of the Schmidt decomposition, there exist examples in which the decomposition in the new basis also has Schmidt form, i.e.\ for which we have:
\eqn{ \label{eqn:basis1_perfect}
\ket{\Psi}_{\mathcal{F}\mathcal{S}} =  \zum{n}{} \, \beta_{n} \, \ket{Fy_n}_{\mathcal{F}} \ket{y_n}_{\mathcal{S}} \, .
}
It follows from \tbf{RQM:5} that $F_Y$ can be interpreted as a pointer variable of $\mathcal{F}$ for the quantity $Y$ of $\mathcal{S}$. Thus the same physical state $\ket{\Psi}_{\mathcal{F}\mathcal{S}}$ allows both $F_X$ to be a pointer variable for $X$, and $F_Y$ to be a pointer variable for $Y$, even though these represent mutually incompatible physical variables.

As an explicit example, consider the case where $\mathcal{F}$ and $\mathcal{S}$ are both spin-half particles and their joint state is the spin-singlet,
\eqn{
\ket{\Psi}_{\mathcal{F}\mathcal{S}} &=& \frac{1}{\sqrt{2}} \left( \ket{z^{+} z^{-}}- \ket{z^{-} z^{+} } \right) \, \\
&=& \frac{1}{\sqrt{2}} \left( \ket{x^{+} x^{-} } - \ket{x^{-} x^{+}} \right) \, ,
}
which is evidently maximally entangled in the spin-z basis $\{ \ket{z^{+}},\,  \ket{z^{-}} \}$ as well as in the spin-x basis $\{ \ket{x^{+}}, \, \ket{x^{-}} \}$ of both particles, which correspond to physical variables.

Returning to the general case, note that there exists a measurement $M$ which $\mathcal{W}$ can perform on the $\mathcal{F}$-$\mathcal{S}$ system that will establish the existence of these correlations. As Rovelli points out (using our own notation):

\begin{quote}
    [T]here is an operator $M$ on the Hilbert space of the $\mathcal{F}$-$\mathcal{S}$ system whose physical interpretation is ``Is the pointer correctly correlated to $X$?" If $\mathcal{W}$ measures $M$, then the outcome of this measurement would be yes with certainty [\dots] [T]he eigenvalue 1 [of $M$] means ``yes, the hand of $\mathcal{F}$ indicates the correct state of $\mathcal{S}$” and the eigenvalue 0 means ``no, the hand of $\mathcal{F}$ does not indicate the correct state of $\mathcal{S}$". [\dots] Thus, it is meaningful to say that, according to the $\mathcal{W}$ description of the events, $\mathcal{F}$ ``knows" the quantity $X$ of $\mathcal{S}$, or that he ``has measured" the quantity $X$ of $\mathcal{S}$, and the pointer variable [$F_X$] embodies the information. --- \cite{ROVELLI_97}, (p.8).
\end{quote}

In our present scenario, it is sufficient to note that Rovelli defines $M$ such that
\eqn{
M \ket{Fx_i} \ket{x_j} := \delta_{ij} \, \ket{Fx_i} \ket{x_j} \,
}
and hence such that $\ket{Fx_i} \ket{x_j}$ is an eigenstate of $M$ with eigenvalue `1'. However, due to the ambiguity of the Schmidt decomposition, one could replace $X$ with $Y$, and $F_X$ with $F_Y$ everywhere in the preceding quote by Rovelli and it would still hold true. The same logic would then compel us to define $M$ using the basis $\{ \ket{Fy_{m}}\ket{y_x} \}$ as:
\eqn{
M \ket{Fy_m} \ket{y_n} := \delta_{mn} \, \ket{Fy_m} \ket{y_n} \,.
}
By measuring $M$ on the state $\ket{\Psi}_{\mathcal{F}\mathcal{S}}$, according to \tbf{RQM:5}, we can assert that $\mathcal{F}$ has measured $\mathcal{S}$ relative to $\mathcal{W}$ -- the problem is that, given the ambiguity in the definition of $M$, \tit{we cannot say which physical variables were involved in the measurement}. 

Since both bases represent physical variables, \tbf{RQM:5} implies that it would be \tit{ad-hoc} to insist that we should use just one definition of $M$ and not the other -- for on what grounds do we decide which one is correct? Yet if we suppose that \tit{both} definitions are somehow simultaneously valid, this would force us to conclude that $\mathcal{F}$ has measured $\mathcal{S}$ simultaneously in incompatible bases relative to $\mathcal{W}$. This is horn (i) of our dilemma. 

To evade this, one could argue that \tbf{RQM:5} was only meant to apply to cases in which the Schmidt decomposition is non-degenerate, hence unique. In the case of degeneracy, we then seem compelled to say that although (relative to $\mathcal{W}$) a measurement has definitely occurred between $\mathcal{F}$ and $\mathcal{S}$ (since this can be checked by measuring $M$), there must be no fact relative to $\mathcal{W}$ about which physical variables were actually measured. 

Even if we grant that such an absurd-sounding statement makes sense, it completely breaks the analogy with relativity, since one clearly cannot have a relativistic relation between two systems without there being some fact of the matter about which physical quantities are thereby related! This is horn (ii) of our dilemma. $\Box$

\subsection{No-go theorem 2 \label{sec:nogo2}}

We next turn to the question of how observers are allowed to disagree about physical states in relativity versus in RQM. In classical relativity the relevant scenario involves two observers $\mathcal{W}$ and $\mathcal{F}$ momentarily in constant motion relative to each other, such that the observers assign different respective velocities $v^{(\mathcal{W})}$ and $v^{(\mathcal{F})}$ to a third system $\mathcal{S}$. In RQM the relevant situation occurs when two observers $\mathcal{W}$ and $\mathcal{F}$ momentarily have different information about a system $\mathcal{S}$ and therefore assign it different states, say $\rho^{(\mathcal{W})}$ and $\rho^{(\mathcal{F})}$ respectively. 

In the classical scenario we can reconcile the different velocity assignments by taking into account the observers' different relative velocities to $\mathcal{S}$, which entails a \tit{mapping} (via the Lorentz transform) between $v^{(\mathcal{W})}$ and $v^{(\mathcal{F})}$. This mapping allows each observer to deduce from their own observations what the velocity of $\mathcal{S}$ should be relative to the other observer. 

In RQM, of course, we cannot expect a deterministic mapping between the state assignments $\rho^{(\mathcal{W})}$ and $\rho^{(\mathcal{F})}$, because the specific outcome of $\mathcal{F}$'s measurement is not a fact relative to $\mathcal{W}$. Nevertheless, one might expect that a weaker non-trivial mathematical constraint holds between $\rho^{(\mathcal{W})}$ and $\rho^{(\mathcal{F})}$. 

Note that for the time being we are not considering the possible ways that two observers might reach agreement, as they might achieve in relativity by accelerating into the same reference frame, and in RQM by measuring one another (we will address that scenario in Sec.\ \ref{sec:nogo4}). Here we are merely pointing out that when observers disagree in classical relativity, their disagreement is still tightly constrained by a strict mathematical mapping, and we would like to know whether an analogous mapping between discordant perspectives can be defined in RQM.

We note that in classical relativity, contrary to \tbf{RQM:4}, it is meaningful to compare two observers' accounts without explicitly invoking further observers. Therefore, in order to proceed, we need to grant at least the following assumption:\\

(i) We can meaningfully define constraints between $\mathcal{W}$'s and $\mathcal{F}$'s state assignments.\\

We will not discuss here whether (i) can be made compatible with \tbf{RQM:4}, that is left to Sec. \ref{sec:nogo3}. For the moment we just assume that it can be compatible, since otherwise the analogy to classical relativity would already be lost.

Supposing the state relative to $\mathcal{W}$ to be a pure entangled state $\ket{\Psi}_{\mathcal{F}\mathcal{S}}$ of the form \eqref{eqn:basis0}, we now ask: according to RQM what are the possible states $\ket{\psi}_{\mathcal{S}}$ that $\mathcal{S}$ could have relative to $\mathcal{F}$? Arguably, the valid states should satisfy the following conditions:\\

(ii) Any valid state assignment $\ket{\psi}_{\mathcal{S}}$ by $\mathcal{F}$ can always be verified by $\mathcal{W}$. That is, there must exists a `pointer basis' of $\mathcal{F}$ such that, if $\mathcal{W}$ were to measure in this basis and condition on the outcome, there would be a nonzero probability of updating the state of $\mathcal{S}$ relative to $\mathcal{W}$ to $\ket{\psi}_{\mathcal{S}}$.\\
(iii) Conversely, any assignment $\ket{\psi}_{\mathcal{S}}$ by $\mathcal{F}$ which can be verified by $\mathcal{W}$ (in the above sense) must be a valid possible assignment for $\mathcal{F}$.\\

Finally, all of this would be pointless if it did not actually constrain the possible states, so we would like to have:\\

(iv) The set of valid states for $\mathcal{F}$ are, in general, a non-trivial subset of the pure states of $\mathcal{S}$.\\

It turns out that if we grant (i)-(iii), then $\mathcal{W}$'s and $\mathcal{F}$'s state assignments do not constrain each other in any interesting way at all. Specifically, we have the following dilemma:\\

\begin{quote}
    \tbf{Dilemma:} The set of assumptions (i)-(iii) are together incompatible with (iv). Specifically, given that $\mathcal{W}$ assigns an entangled state $\ket{\Psi}_{\mathcal{F}\mathcal{S}}$ of the form \eqref{eqn:basis0}, and assuming the coefficients $\alpha_i$ are all nonzero, then \tit{every} pure state in the Hilbert space of $\mathcal{S}$ is a possible state relative to $\mathcal{F}$.
\end{quote}

This is a well-known result whose proof we shall not repeat here. To put it briefly, if the support of the reduced state $\rho_{\mathcal{S}} := \trm{tr}_{\mathcal{F}}( \ketbra{\Psi}{\Psi})$ spans the Hilbert space of $\mathcal{S}$, then it is possible to \tit{steer} $\mathcal{S}$ into an arbitrary pure state with nonzero probability by measuring $\mathcal{F}$ and conditioning on the outcome\footnote{The first known derivation of this result is by Schr\"{o}dinger; it has since been rediscovered and expanded upon by subsequent authors; see Refs.\ \cite{STEERING,SHJW} and citations therein.}.

One might try to escape this conclusion by insisting that some of the $\alpha_i$ can be exactly zero; however we think this is ineffective for two reasons: (1) it is an idealization to assume that two systems could be coupled in certain variables but strictly isolated in others -- all variables within a system `talk' to each other; (2) even allowing such cases, their occurrence would not be typical, seeing as they constitute a set of measure zero in the space of physically possible states.

To conclude our point: when two observers are in a situation where they disagree about the state of a system in RQM, the state relative to one observer \tit{places no non-trivial constraints on the state relative to the other observer}, in stark contradistinction to disagreements about velocity and other classical quantities in relativity.

\subsection{Concluding remarks on the relativistic analogy}

We have seen in this section that RQM's rejection of the absoluteness of relations via its rejection of (A1-A3) raises a question: is there anything left of the classical relativistic intuitions that RQM is supposed to preserve? Rovelli and Laudisa write that RQM is merely a weakening of realism ``in a direction similar to what happened with Galilean or Einstein’s relativity, [\dots] But is a more radical step in this direction" \cite{SEP_RQM}. On the contrary, we have shown that RQM implies a radical rejection of all the most essential intuitive features of classical relativistic relations\footnote{We acknowledge that Dorato already made this general point in Ref. \cite{RQM_DORATO}; we have just shored it up with mathematical arguments.}. The mere fact that states are indexed to observers does not by itself warrant the descriptor `relational'. Far from having de-mystified quantum mechanics by appealing to relations, RQM has merely mystified the concept of a `relation'. 

We conclude that the only thing `relational' about RQM is its adherence to the principle of relative facts. By that measure, QBism should also be called `relational'. Perhaps more accurate would be to rename RQM the `Relative-facts interpretation of quantum mechanics', but this would not appeal to anyone who hoped for a passing resemblance to relativity theory.

\section{Challenges to the objectivity of RQM \label{sec:not_objective}}

There is another feature of RQM besides its appeal to classical relativistic intuitions that makes it attractive to many: its claim to \tit{objectivity}. Of course, in light of the preceding section `objective' must mean something other than `absolute', so clarification is needed. Rovelli recognized this in 1996 when he wrote

\begin{quote}
[W]e are forced to accept the result that there is no ``objective," or more precisely ``observer-independent," meaning to the ascription of a property to a system.
\end{quote}

Despite the observer-dependence of properties and quantum events generally, there are three ways in which RQM still stakes a claim to a weaker, `relational' notion of objectivity.

The first and most obvious is RQM's commitment to \tit{physicalism}, namely, the idea that fundamental reality is comprised of nothing but physical systems and the laws governing them \cite{SEP_PHYSICALISM}. Hence in RQM, the concepts of a broadly mental character like `experience' and `consciousness' are at best emergent phenomena, if they can be said to exist at all. It follows that relations in RQM are strictly indexed to \tit{objects}, rather than to (thinking and feeling) \tit{subjects}. Since this is a deep philosophical commitment not amenable to physicists' blunt instruments like no-go theorems, we shall leave its discussion to the companion paper, Ref.\ \cite{QBISM_COMPARISON}, in which it will be better illuminated by contrasting it to QBism's radical subjectivism. For present purposes, we focus on the two claims \tbf{RQM:1: any system can be an observer} and \tbf{RQM:5: any physical correlation is a measurement}, which may be thought of as corollaries of RQM's commitment to physicalism.

A second form of relational objectivity in RQM comes from the claim that the relation between any pair of systems (i.e.\ the state of one relative to the other) is fully determined by the past histories of those systems alone, as expressed in \tbf{RQM:3: relations are intrinsic}. This bestows a certain `objectivity' on any relation between two systems, in the sense that events extraneous to the systems' joint past measurement history cannot have any bearing on the relation between them.

The third form of relational objectivity comes from the claim that agreement about a quantum event necessarily occurs between two observers' accounts when one observer measures another observer's pointer variable, as expressed in \tbf{RQM:6: shared facts}. A measurement event can therefore attain a kind of `partial objectivity', in the sense that its value can be verified by multiple observers, if they all measure in the basis appropriate to the `pointer variable' in which the value was recorded.

In the remaining subsections we will present three distinct arguments challenging RQM's claim to objectivity. Specifically, in Sec.\ \ref{sec:nogo3} we examine arguments in the primary literature that purport to derive \tbf{RQM:6} from the quantum formalism, and conclude that none of these arguments achieve this goal. In Sec.\ \ref{sec:nogo4} we argue that \tbf{RQM:6} is actually in conflict with the other principles of RQM, while in Sec.\ \ref{sec:nogo5} we show a conflict between \tbf{RQM:1}, \tbf{RQM:3} and \tbf{RQM:5}.

\subsection{No-go theorem 3 \label{sec:nogo3}}

Reading the primary literature on RQM, it is difficult not to be struck by what appears to be a manifest tension between two of its principles. On the one hand, it is frequently claimed that when one observer measures another observer, quantum theory guarantees consistency or agreement between their accounts of events (\tbf{RQM:6}). On the other hand, it is just as frequently emphasized that it is only meaningful to compare two observers' accounts by encompassing them within a single account relative to just one observer (\tbf{RQM:4}).

Although the issue has been raised many times before in the secondary literature (see eg.\  Refs.\cite{RQM_BROWN, RQM_RUYANT, RQM_VANF, RQM_MOS}), the primary literature steadfastly maintains that there is no conflict between these two principles. It is important to achieve clarity about this issue, because it has serious implications for the ontology of RQM. Therefore, in the interests of progress, we shall revisit the issue once more by formulating it as a dilemma:

\begin{quote}
    \tbf{Dilemma:} RQM cannot consistently maintain both the principle of \tbf{RQM:6: shared facts}, and the principle of \tbf{RQM:4: comparisons are relative to one observer}. Rejecting one or the other leads to the following two horns:\\
(i) If RQM rejects \tbf{RQM:6}, then it either implies \tit{solipsism}, or else an \tit{ontology of island universes} (these terms will be defined at the end of this section).\\
(ii) If RQM instead rejects \tbf{RQM:4}, it becomes vulnerable to our next no-go theorem, in Sec. \ref{sec:nogo4}.
\end{quote}

\tit{Proof:} To begin with, note that in RQM an `observer' can appear in one of two distinct roles: either as the `context' or `frame' relative to which other systems are assigned states, or as a system relative to another observer's context. To avoid confusion, it is important to distinguish between these two roles.

According to \tbf{RQM:4}, any meaningful claims about physical quantities or states of affairs must be referred to just one observer; we shall call the observer in this role the \tit{frame-observer}. Since observers in general are just physical systems, we can describe other observers as physical systems relative to the frame-observer. By definition, these `observed observers' do not themselves occupy the role of frame-observer, so we shall say that they stand in the role of \tit{secondary observers}. Thus, whenever RQM talks about the \tit{account} or \tit{perspective} of an observer, we may interpret this as referring to the account or perspective of that observer \tit{when they are the frame-observer}. 

Our main argument is simply that there is no way to meaningfully compare the accounts of two frame-observers. It is simple enough to see why: only one observer can be the frame-observer at a time. Any observer can of course stand in the role of the frame-observer; the particular choice is a matter of convention only. However, to carry out a consistent analysis of any statement, we must select only one observer to play this role. A comparison requires two -- the conclusion follows.

More concretely, consider two systems $\mathcal{W}$ and $\mathcal{F}$, observing each other as well as some third system $\mathcal{S}$. We can of course choose either $\mathcal{W}$ or $\mathcal{F}$ to serve as frame-observer, and in general each will have a different account of events. Specifically, $\mathcal{W}$ and $\mathcal{F}$ can in general assign different quantum states to $\mathcal{S}$, and may have different histories of measurements of $\mathcal{S}$ (i.e.\ they can have different perspectives). Now suppose we have before us a description of $\mathcal{W}$'s account, and a description of $\mathcal{F}$'s account -- laid out `side by side' in a view from nowhere, so to speak -- and we would like to know: \tit{are these accounts mutually consistent?} 

On the one hand, \tbf{RQM:6} requires that this question be well-posed, for otherwise there would be no way to assert that two observer's accounts are in agreement. On the other hand, according to \tbf{RQM:4}, this is not a well-posed question, because there is no `view from nowhere'. The conflict becomes all the more vexing when one considers statements like Rovelli's assertion in Ref.\ \cite{ROVELLI_96} that ``different observers may give different accounts of the same sequence of events". On what basis can it be asserted that two observers' accounts are different, if there is no observer capable of verifying the difference by a measurement? Indeed, on RQM's own account, it appears that no such observer can exist, for any attempt to check consistency between two observers by physically measuring them is mandated to result in agreement. This establishes the dilemma. $\Box$

Before we turn to the consequences of this dilemma, it is important to understand why the primary literature has failed to acknowledge the conflict between \tbf{RQM:6} and \tbf{RQM:4}. In the remainder of this section we argue that this is due to a tendency to conflate the two roles of an observer, thereby leaving it ambiguous as to whether the observer is a frame-observer or merely a secondary observer. This creates what we call a `loose-frame loophole', in which a claim is presented as though it establishes consistency between two frame-observer's accounts, thus apparently supporting \tbf{RQM:6}, when in fact upon examination the claim only implies internal consistency within a single frame-observer's account. This ultimately undermines the apparent textual support for \tbf{RQM:6} in the literature.

\subsubsection{The loose frame loophole}

In the primary literature it is usually claimed that the necessity of agreement between distinct observers' accounts is derivable from the formalism of quantum theory. The prototypical argument, following Rovelli's 1996 paper, runs as follows: consider $\mathcal{W}$'s account of the observer $\mathcal{F}$ measuring a quantity $X$ of the system $\mathcal{S}$. Rovelli writes (using our own labels; emphasis in original):

\begin{quote}
[T]here is a consistency condition to be fulfilled, which is the following: if $\mathcal{W}$ knows that $\mathcal{F}$ has measured $X$, and then she [$\mathcal{W}$] measures $X$, and then she measures what $\mathcal{F}$ has obtained in measuring $X$, consistency requires that the results obtained by $\mathcal{W}$ about the variable $X$ and the pointer [of $\mathcal{F}$] are correlated [\dots] From the point of view of the $\mathcal{W}$ description: 
\tit{The fact that the pointer variable in $\mathcal{F}$ has information about $\mathcal{S}$ (has measured $X$) is expressed by the existence of a correlation between the $X$ variable of $\mathcal{S}$ and the pointer variable of $\mathcal{F}$. The existence of this correlation is a measurable property of the $\mathcal{F}$-$\mathcal{S}$ state.} --- \cite{ROVELLI_96}, (p.1652).
\end{quote}

What is established by the above argument is only that there is a correlation between $\mathcal{W}$'s account of the value of $X$, and $\mathcal{W}$'s account of the value of $\mathcal{F}$'s pointer variable. RQM does not furnish us with any principle that would allow us to associate ``$\mathcal{W}$'s account of $\mathcal{F}$'s pointer variable" with ``$\mathcal{F}$'s account of $\mathcal{F}$'s pointer variable". We have no formal means of asserting that $\mathcal{W}$'s account of the reading of $\mathcal{F}$'s instruments has any bearing on $\mathcal{F}$'s own account of their instrument readings when $\mathcal{F}$ is the frame-observer.

As much as it seems reasonable to suppose that two frame-observer's accounts should agree when they check the readings of each other's instruments -- as is necessary for \tbf{RQM:6} to be upheld -- we cannot rely on a mere supposition; rigor demands that it either be postulated explicitly, or else derived from more basic principles. RQM does not postulate it, and as the remainder of this section shows, also fails to derive it.

We next consider a rehashing of Rovelli's argument by Laudisa and Rovelli \cite{SEP_RQM} (our labeling):

\begin{quote}
Prima facie, RQM may seem to imply a form of perspective solipsism, as the values of variables realized in the perspective of some system $\mathcal{F}$ are not necessarily the same as those realized with respect to another system $\mathcal{W}$. This is however not the case, as follows directly from quantum theory itself. The key is to observe that any physical comparison is itself a quantum interaction. Suppose the variable $X$ of $\mathcal{S}$ is measured by $\mathcal{F}$ and stored into the variable $F_X$ of $\mathcal{F}$. This means that the interaction has created a correlation between $X$ and $F_X$. In turn, this means that a third system measuring $X$ and $F_X$ will certainly find consistent values. That is: the perspectives of $\mathcal{F}$ and $\mathcal{W}$ agree on this regard, and this can be checked in a physical interaction. --- \cite{SEP_RQM}
\end{quote}

The only significant difference to the original argument is that it is now unclear who the frame-observer is supposed to be. The phrasing ``the interaction has created a correlation between $X$ and $F_X$" is ambiguous: relative to \tit{which system} has the interaction taken place? Relative to \tit{which system} is there a correlation? There is a loose frame here.

The simplest assumption is that the ambiguous statements are meant to refer to the system $\mathcal{W}$ serving as frame-observer. (We note that one could also use some other system $\mathcal{C}$ for this role, and our argument would reach the same conclusions). Here is the relevant part of the text again, with our clarifications added in bold: 

\begin{quote}
Suppose the variable $X$ of $\mathcal{S}$ is measured by $\mathcal{F}$ \tbf{(relative to $\mathcal{W}$'s account)} and stored into the variable $F_X$ of $\mathcal{F}$ \tbf{(relative to $\mathcal{W}$)}. This means that the interaction has created a correlation between $X$ and $F_X$ \tbf{(relative to $\mathcal{W}$)}. In turn, this means that a third system \tbf{(i.e.\ $\mathcal{W}$ )} measuring $X$ and $F_X$ will certainly find consistent values. [\dots]
\end{quote}

And now that we have removed the ambiguity, it is evident that the concluding line does not logically follow:

\begin{quote}
That is: the perspectives of $\mathcal{W}$ and $\mathcal{F}$ agree on this regard, and this can be checked in a physical interaction.
\end{quote}

A `physical interaction' relative to whom? Presumably, relative to $\mathcal{W}$. But then, since all preceding statements also took $\mathcal{W}$ as the frame-observer, the most we can conclude is that there is consistency between $\mathcal{W}$'s account of $X$ and $\mathcal{W}$'s account of $F_X$, none of which has any logical implications for $\mathcal{F}$'s own perspective \tit{when $\mathcal{F}$ is the frame-observer}.

Finally, let us consider a more subtle instance of the loophole that appears in recent work of Di Biagio and Rovelli \cite{RQM_STABLE_FACTS}. The authors contemplate a scenario in which system $\mathcal{F}$ has measured a variable $L_\mathcal{S}$ of $\mathcal{S}$ and obtained one of a set of mutually exclusive outcomes $a^{(\mathcal{F})}_i (i=1,\dots,N)$, where the superscript $(\mathcal{F})$ reminds us that these represent facts relative to $\mathcal{F}$ (we shall adhere to the authors' notation). We are then asked to contemplate a possible outcome $b^{(\mathcal{W})}$ of a subsequent measurement that $\mathcal{W}$ performs on an unspecified system\footnote{A subscripted index on $b^{(\mathcal{W})}$ would be desirable to maintain consistency of the notation, but we will scrupulously follow the notation of the original authors.}, where again the superscript reminds us that $b^{(\mathcal{W})}$ is a fact relative to $\mathcal{W}$. The authors then propose to define the outcomes $a^{(\mathcal{F})}_i$ as being \tit{stable facts} relative to the observer $\mathcal{W}$ if and only if the probabilities $P(b^{(\mathcal{W})})$ satisfy the equation:
\eqn{ \label{eqn:stable}
P(b^{(\mathcal{W})}) = \zum{i}{N} \, P(b^{(\mathcal{W})}|a^{(\mathcal{F})}_i) \, P(a^{(\mathcal{F})}_i) \, ,
}
and they observe that this condition does not hold in general. The remainder of the paper discusses its implications and the conditions under which it might hold.

However, we are prompted to ask: relative to which observer are the conditional probabilities $P(b^{(\mathcal{W})}|a^{(\mathcal{F})}_i)$ defined? Given the central role of Eq.\ \eqref{eqn:stable} in their arguments, it is surprising that there is only \tit{one} place in the text that indicates the frame to which the probability is referred. The authors write (our emphasis):

\begin{quote}
[T]he probabilities for facts of the $\mathcal{S}$-$\mathcal{F}$ system \tit{relative to} $\mathcal{W}$ can indeed be computed from an entangled state of the form [\dots], where $a_i$ are values of $L_\mathcal{S}$ and $Fa_i$ are values of $\mathcal{F}$'s `pointer variable' $L_\mathcal{F}$. --- \cite{RQM_STABLE_FACTS}, (p.3).
\end{quote}

This clarifies that the probabilities are meant to be defined relative to system $\mathcal{W}$. In connection with this, note that in the above quote the authors have -- without remark -- dropped the $(\mathcal{F})$ superscript from the $a_i$. Why did they do so?

Presumably because, as we have established, the $a_i$ in the sentence just quoted do not refer to facts relative to $\mathcal{F}$, but rather to facts relative to $\mathcal{W}$, and so should be indexed as $a^{(\mathcal{W})}_i$. Once this is made explicit, a problem arises: how can these probabilities, which refer to $a^{(\mathcal{W})}_i$, be compared against the condition \eqref{eqn:stable}, which refers instead to $a^{(\mathcal{F})}_i$? Yet the authors claim that such a comparison can be made. The loose frame loophole is at work.

The only way for the comparison to be legitimate is to postulate an identification, such as $a^{(\mathcal{W})}_i \equiv a^{(\mathcal{F})}_i$. But nowhere in the text is it admitted that such an additional postulate is required. The closest the authors come to this admission is when they discuss a situation where the $a^{(\mathcal{F})}_i$ are stable relative to $\mathcal{W}$, and they write: 

\begin{quote}
The observer [$\mathcal{W}$] might say ``$L_{\mathcal{S}}$ has been measured," and assume that the pointer of the apparatus moved one way or the other. In the mathematical formalism, $\mathcal{W}$ can assume that ``$\mathcal{S}$'s wavefunction has collapsed." Note however that neither the value of $L_{\mathcal{S}}$ nor that of $L_{\mathcal{F}}$ is a fact for $\mathcal{W}$ at this stage. Stability simply allows $\mathcal{W}$ to ``de-label" facts relative to $\mathcal{F}$. --- \cite{RQM_STABLE_FACTS} (p.6).
\end{quote}

But is this `de-labeling' intended to be a statement about the \tit{ontological status} of the facts $a^{(\mathcal{W})}_i$ and $a^{(\mathcal{F})}_i$? Are we to read this paragraph as making the \tit{ontological postulate} that we can identify these two relative facts as being a \tit{single} relative fact, relative to both $\mathcal{W}$ and $\mathcal{F}$, i.e.\ $a^{(\mathcal{W})}_i \equiv a^{(\mathcal{F})}_i \equiv a_i$? The authors do not answer this question directly, but their phrasing in the subsequent paragraph suggests that their answer is negative. They write (emphasis in original):

\begin{quote}
[\dots] [B]ased on the value of [the pointer variable] $L_{\mathcal{F}}$, the experimenter $\mathcal{W}$ can update the state for $\mathcal{S}$. The experimenter can reason \tit{as if} $L_{\mathcal{S}}$ took the value that she read on the apparatus' pointer variable [$L_{\mathcal{F}}$].
\end{quote}

In RQM, of course, $\mathcal{W}$ is nothing but a physical system, hence we should take the authors' reference to an ``experimenter" who ``can reason as if \dots" with a large grain of salt (this is not QBism after all). A more consistent reading of this passage is that the identification $a^{(\mathcal{W})}_i \equiv a^{(\mathcal{F})}_i$ may be assumed for purely \tit{methodological} purposes, and calculation may be carried out ``as if" it were the case. The implication, of course, is that \tit{it isn't really the case}, hence no actual identification at the ontological level is being proposed.

\subsubsection{An ontology without shared facts?}

Our preceding analysis suggests that despite surface appearances, the primary literature does not in fact support \tbf{RQM:6: shared facts}, and the dilemma posed at the beginning of this section leads us to reject it. In this section we ask: what would the ontology of RQM look like in the absence of \tbf{RQM:6}?

The answer hinges on whether and how we conceive of the totality of all facts, relative to multiple frame-observers. We have two options: we can allow that it is meaningful to contemplate the totality of `relative facts' for all observers, provided that we abstain from sharing or comparing facts relative to different observers, leaving them to exist as it were `in parallel', or else we can take a stronger line and insist that it is meaningless to contemplate any facts beyond those which are relative to a single designated frame-observer.

In the primary literature, quotations abound that appear to support both views, often within the same paper. For instance, in Smerlak \& Rovelli one finds that `` reality admits one description per (observing) system,
each being internally consistent", which seems to suggest that it is meaningful to speak of reality \tit{in toto} as comprising a collection of descriptions, one per system; yet in the same paper we find that ``the collection of all properties relative to all systems [\dots] is assumed not to make sense". Charitably, we can interpret this to mean that both possibilities remain open.

Out of the two options, the second is much less appealing, as it effectively implies that there is only a single frame-observer who is the ultimate referent for what is `real'. One can perhaps hypothesize and speculate about reality relative to other observers, but per definition the reality of others is purely a convenient fiction: all relative facts are relative to just one observer. This meets one of the common definitions of \tit{solipsism}, namely, that no meaning can be given to reality except as indexed to a single observer \cite{IEP_SOLIPSISM}. Given that the primary literature claims to explicitly reject solipsism, this option seems an unlikely fit for RQM. 

The first option still allows for the in-principle existence of multiple observers, but requires that their sets of relative facts be incomparable. This is not solipsism in any usual sense: each observer's private world is still real and external to themselves, and each observer considers the others to really exist. Rather, we might call it an ontology of `island universes', inspired by this passage by Aldous Huxley:

\begin{quote}
We live together, we act on, and react to, one another; but always and in all circumstances we are by ourselves. The martyrs go hand in hand into the arena; they are crucified alone. [\dots] We can pool information about experiences, but never the experiences themselves. From family to nation, every human group is a society of island universes. --- \tit{The Doors of Perception} \cite{HUXLEY}, (p.12).
\end{quote}

Of course, since any system can be an observer, the island universes in RQM are not just confined to human perspectives as in Huxley's vision: rather, \tit{every fundamental particle in existence would be associated to a unique `island universe'}. This proliferation of disjoint universes is not motivated by observations, nor does it serve any explanatory purpose; it therefore seems to be (paraphrasing Wheeler) ontological `excess baggage'. Moreover, it is difficult to see what `objectivity' could possibly mean in such a universe. 

\subsection{No-go theorem 4 \label{sec:nogo4} }

In light of the preceding, it is tempting to go against the grain of the primary literature and take the second horn of the dilemma outlined in Sec.\ \ref{sec:nogo3}, that is, to reject \tbf{RQM:4} in favour of postulating \tbf{RQM:6}. Note that in order to do so, we shall have to adopt a standpoint in which the accounts of $\mathcal{W}$ and $\mathcal{F}$ can be considered `side by side' as it were, effectively appealing to a `view from nowhere'. Nevertheless, it is instructive to ask whether making this concession might allow for a consistent ontology more appealing than the ones considered in the preceding section. 

Perhaps contrary to expectations, we find that even after dropping \tbf{RQM:4} there is still no straightforward way to postulate \tbf{RQM:6} between frame-observers in a manner consistent with the other principles of RQM. 

Let us assume that in $\mathcal{W}$'s account, $\mathcal{F}$ has measured $\mathcal{S}$ in basis $\{ \ket{x_i} \}$, and recorded the outcome in their pointer variable $\{ \ket{Fx_{i}} \}$; call this ``Event 1". Recall that, according to RQM, just after Event 1 the state of the joint system $\mathcal{F}$-$\mathcal{S}$ in $\mathcal{W}$'s account has the form: 
\eqn{ \label{eqn:basis0_again}
\ket{\Psi}_{\mathcal{F}\mathcal{S}} = \zum{i}{}  \, \alpha_{i} \, \ket{Fx_{i}}_{\mathcal{F}} \ket{x_i}_{\mathcal{S}} \, .
}
Let us now ask what happens when $\mathcal{W}$ measures $\mathcal{F}$'s apparatus pointer, i.e.\ the situation in which \tbf{RQM:6} would be expected to hold. For the sake of generality, suppose that at some time after Event 1, $\mathcal{W}$ measures $\mathcal{F}$-$\mathcal{S}$ in some basis, where for the sake of argument we assume that the particular basis is unconstrained. Namely, we assert:\\

\begin{quote}
    \tbf{P1.} $\mathcal{W}$ can measure $\mathcal{F}$-$\mathcal{S}$ in any basis at Event 2, independently of which basis $\mathcal{F}$ measured $\mathcal{S}$ at Event 1.
\end{quote}

There is some ambiguity here, because if $\mathcal{W}$ measures in a basis incompatible with $\mathcal{F}$, it is not clear whether the outcome obtained by $\mathcal{F}$ at Event 1 should constrain the outcome obtained by $\mathcal{W}$ at Event 2. Fortunately we shall not need to answer that question in order to make our argument; what matters for our purposes is not the precise \tit{outcome} of the measurement, but rather which \tit{basis} it was performed in. Specifically, assuming as is our habit that the Schmidt decomposition is degenerate and hence non-unique, let $F_Y$ and $Y$ be another set of perfectly correlated physical variables between $\mathcal{F}$ and $\mathcal{S}$, corresponding to the basis $\{ \ket{Fy_m} \ket{y_n} \}$. Then \tbf{RQM:6} implies the following:

\begin{quote}
    \tbf{P2:} Suppose $\mathcal{W}$ measures $\mathcal{F}$-$\mathcal{S}$ in the $\{ \ket{Fy_m} \ket{y_n} \}$ basis and obtains some outcome, updating the state relative to $\mathcal{W}$ to one of the states in $\{ \ket{Fy_n}\ket{y_n} \}$ just after Event 2. Then we can interpret this state as indicating that `$\mathcal{F}$ measured $\mathcal{S}$ in the $\{ \ket{y_n} \}$ basis and obtained one of the outcomes in the set $\{ y_n \}$ at Event 1'.
\end{quote}

Now the contradiction is manifest: for if $\mathcal{F}$ measured $\mathcal{S}$ in the $\{ \ket{y_n} \}$ basis, then she could not have measured it in the $\{ \ket{x_i} \}$ basis as originally supposed. According to \tbf{P1}, we cannot enforce consistency by making $\mathcal{W}$'s measurement at Event 2 depend on the basis of $\mathcal{F}$'s measurement at Event 1. Therefore our only option is to do the reverse: the basis of $\mathcal{F}$'s measurement at Event 1 must depend on the basis of $\mathcal{W}$'s measurement at Event 2. This violates \tbf{RQM:3} (relations are intrinsic) because the latter is in the causal future of Event 1, i.e.\ it is not part of the perspectives of either $\mathcal{F}$ or $\mathcal{S}$ at Event 1. Thus we have a trilemma:

\begin{quote}
    \tbf{Trilemma:} The propositions \tbf{P1} \& \tbf{P2} and the claim \tbf{RQM:3} cannot all be true.
\end{quote}

As a concrete example, consider the spin-singlet state from Sec.\ \ref{sec:nogo1} and suppose that $\mathcal{F}$ measured $\mathcal{S}$ in the spin-z basis at Event 1 and assigned it one of the states $\{ \ket{z^{+}},\ket{z^{-}} \}$. By \tbf{P1}, $\mathcal{W}$ is free to measure in the spin-x basis at Event 2 and assign $\mathcal{F}$-$\mathcal{S}$ one of the states $\{ \ket{x^{+}}\ket{x^{+}}, \, \ket{x^{-}}\ket{x^{-}} \}$. But then \tbf{P2} asserts that $\mathcal{F}$ must have measured $\mathcal{S}$ in the spin-x basis at Event 1 an obtained one of the states $\{ \ket{x^{+}}, \ket{x^{-}} \}$ -- a contradiction.

Let us now consider some ways of avoiding the trilemma. One might object that \tbf{RQM:6} only applies when $\mathcal{W}$ measures in the same basis as $\mathcal{F}$, and hence that we should reject \tbf{P2}. However, this objection is cut off by \tbf{RQM:5}, which asserts that a perfect correlation between \tit{any} pair of physical variables counts as a measurement interaction between those variables, with one variable representing the apparatus pointer that indicates the value of the other variable. By that token, we have no physically motivated grounds for treating the $\{ \ket{Fy_m} \ket{y_n} \}$ basis any differently to the $\{ \ket{Fx_i} \ket{x_j} \}$ basis.

Suppose we try to resolve this dilemma by rejecting \tbf{P1}. Then consistency requires that the choice of basis of $\mathcal{F}$'s measurement at Event 1 determines the basis of $\mathcal{W}$'s subsequent measurement at Event 2\footnote{Some readers might worry that this violates $\mathcal{W}$'s `freedom of choice', but it is not at all clear that this is a legitimate complaint in RQM; after all, what if $\mathcal{W}$ were just an asteroid, or an electron?}.

The troubling feature of this option is that the variable that specifies the basis of $\mathcal{F}$'s measurement at Event 1 now becomes equivalent to a \tit{hidden variable} for the measurement at Event 2. Note that it carries information relevant for predicting something about $\mathcal{W}$'s quantum event at Event 2 (namely, its basis) and that it possesses a value in the ontology (relative to $\mathcal{F}$) prior to Event 2, such that if this value were accessible to $\mathcal{W}$, it would lie within $\mathcal{W}$'s causal past. Yet it is independent of all of $\mathcal{W}$'s previous quantum events, so it meets our definition of a \tit{hidden variable} and thereby violates \tbf{RQM:2}.

In summary, RQM faces a trilemma, which leads to choosing between the following horns:
\begin{itemize}
    \item(i) Reject \tbf{RQM:3}, implying that relations are not intrinsic to pairs of systems;
    \item(ii) Reject \tbf{P1}, implying that RQM admits a type of hidden variables, contrary to \tbf{RQM:2};
    \item(iii) Reject \tbf{P2}, implying that we must give up either \tbf{RQM:5: any physical correlation is a measurement}, or \tbf{RQM:6: shared facts}.  
\end{itemize}

\subsection{No-go theorem 5 \label{sec:nogo5}}

Our final result reveals a conflict between \tbf{RQM:1: any system can be an observer}, \tbf{RQM:3: relations are intrinsic} and \tbf{RQM:5: any physical correlation is a measurement}. Our argument depends crucially on the assumption that both $\mathcal{F}$ and $\mathcal{S}$ can be spin-half systems; for this reason we call it the \tit{Wigner's little friend} paradox. 

We derive our trilemma from the following simple result:

\begin{quote}
    \tbf{Result:} There can exist perfect correlations between the physical variables of two spin-half particles, which can be altered by performing global rotations of the two-particle system relative to an external set of spatial axes.
\end{quote}

\tit{Proof:} Consider the `triplet' Bell states:
\eqn{
\ket{\Psi^{+}} &:=& \frac{1}{\sqrt{2}} \left( \ket{\uparrow \downarrow} + \ket{\uparrow \downarrow}\right)\, , \nonumber \\
\ket{\Phi^{+}} &:=& \frac{1}{\sqrt{2}} \left( \ket{\uparrow \uparrow}+ \ket{\downarrow \downarrow}\right) \, , \nonumber \\
\ket{\Phi^{-}} &:=& \frac{1}{\sqrt{2}} \left( \ket{\uparrow \uparrow}- \ket{\downarrow \downarrow}\right) \, ,
}
where $\{\ket{\uparrow}, \ket{\downarrow} \}$ is an orthonormal basis representing the spin component along some arbitrary fixed axis. These states have the following properties. First, each one represents perfect correlations (or anti-correlations) between the spins of the two particles along the chosen reference axis, hence between physical variables of the two particles. Secondly, each state represents a \tit{distinct} correlation, as they can be perfectly distinguished from one another by measuring the joint system in the Bell basis, $\{\ket{\Psi^{+}}, \ket{\Psi^{-}}, \ket{\Phi^{+}}, \ket{\Phi^{-}} \}$, which is moreover associated to a physical variable of the joint system (that is, one which `we know how to measure'). Third, the above three states are all $S^2=1$ eigenstates of the total spin operator, which are known to be convertible into one another by \tit{global rotations} of the joint system about the appropriate spatial axes. This establishes the result. $\Box$

How does this create a problem for RQM? To begin, we identify the systems $\mathcal{F}$ and $\mathcal{S}$ as the two spin-half particles, and assume they are assigned one of the triplet Bell states relative to the spatial axes of a third system $\mathcal{W}$. According to \tbf{RQM:5}, if $\mathcal{W}$ measures $\mathcal{F}$-$\mathcal{S}$ in the Bell basis, he can confirm that a physical interaction has occurred between the particles, and by \tbf{RQM:1}, this interaction establishes a relation between the two spin-half particles in which one of them can be said to have `observed' the other.

Note that since there are three states corresponding to physically distinguishable measurement outcomes in the Bell basis, there are correspondingly three different relations that could be established between the two systems; we label them accordingly $R^i$ with $i \in \{\Psi+, \Phi-, \Phi+ \}$.

Finally, note that by rotating the orientation of his spatial axes relative to the two-particle system, $\mathcal{W}$ can arrange to make the state of $\mathcal{F}$-$\mathcal{S}$ (relative to himself) transform into any one of the other triplet Bell states above, thereby changing which of the relations $\{ R^i \}$ obtains relative to himself. Moreover, he can do this without interacting with $\mathcal{F}$-$\mathcal{S}$, i.e.\ without the fact of the rotation of his spatial axes entering into their perspectives. This means that the relation which $\mathcal{W}$ has established to exist between $\mathcal{F}$ and $\mathcal{S}$ cannot be intrinsic to that pair of systems, contradicting \tbf{RQM:3}.

Thus we have a trilemma that forces us to choose among the following horns:
\begin{itemize}
    \item(i) Reject \tbf{RQM:3}, by asserting that relations are not necessarily intrinsic to pairs of systems;
    \item(ii) Reject \tbf{RQM:1}, by asserting that spin-half particles do not count as observers;
    \item(iii) Reject \tbf{RQM:5}, by asserting that relations between systems entail more than just perfect correlation between physical variables.
\end{itemize}

We note that this trilemma does not involve \tbf{RQM:6: shared facts}, so it would still bite us even if we were to take the drastic step of embracing an island-universe (or solipsistic) ontology in response to the previous no-go theorems.

\section{Conclusions}
We have argued that RQM appeals to two main ideas: that it preserves certain classical relativistic intuitions about relations, and that it preserves certain objective properties of relations, in particular the idea that consistency can be established between different observers' accounts.

In Sec.\ \ref{sec:not_relational} we challenged the supposed analogy with classical relativity. In Sec.\ \ref{sec:nogo1} we showed that the relations in RQM cannot always be assigned definite physical quantities, or else relations between physically incompatible quantities can sometimes co-exist. In Sec.\ \ref{sec:nogo2} we argued that there is no non-trivial analog in RQM of the transformation between classical relativistic observers.

In Sec.\ \ref{sec:not_objective} we challenged the `objectivity' of RQM. We argued that RQM's claim to objectivity depends on upholding the three claims \tbf{RQM:1},\tbf{RQM:3} and \tbf{RQM:6}. Subsequently we presented three no-go theorems placing these claims in various conflicts with each other and with the other principles of RQM. 

Specifically, in Sec.\ \ref{sec:nogo3} we revealed a conflict between \tbf{RQM:4: comparisons are relative to one observer} and \tbf{RQM:6: shared facts}, and argued that claims purporting to establish \tbf{RQM:6} in the literature do not succeed because of a \tit{loose frame loophole}: they failed to distinguish between an observer's own account, versus the account of an observer as given by another observer. Furthermore, we argued that giving up \tbf{RQM:6} in favour of \tbf{RQM:4} would imply an undesirable ontology of `island universes'.

In Sec.\ref{sec:nogo4} we showed that even if \tbf{RQM:6} were upheld to enable agreement between different observers, certain situations would imply a disagreement about which \tit{basis} the measurement was performed in, which then leads to conflicts with \tbf{RQM:2}, \tbf{RQM:3}, or \tbf{RQM:5}. 

Finally in Sec.\ref{sec:nogo5} we argued that if spin-half systems can be observers as implied by \tbf{RQM:1}, then there is a special situation in which a conflict arises with \tbf{RQM:3} and \tbf{RQM:5}. This occurs due to a special property of the triplet Bell states of two spins, namely, their ability to sustain perfect correlations in mutually incompatible bases which can be transformed into one another by global rotations.

In light of these challenges, it is not clear that RQM represents an internally coherent ontology for quantum theory. It is certainly possible that coherence might be restored by strategically navigating between the horns of the various multi-lemmas presented here. However, it is unclear whether the resulting ontology, if coherent, would amount to anything more than the doctrine of fact-relativity taken to its logical extreme: an ontology of disconnected island universes, one for each system in existence. If that is indeed the ontology that turns out to be required by internal consistency, then it would seem to do little towards dispelling the non-intuitive features of quantum theory.

Is the situation hopeless? Perhaps not. In quantum foundations, every paradox or no-go theorem can be taken as a guide towards making an improved interpretation, even if that means giving up some previously dearly held ideas. We hope that proponents of RQM will do their utmost to address the challenges presented here. In particular, it would be interesting to see whether a decisive justification could be given for RQM's claim \tbf{RQM:6}, allowing facts to be shared between frame-observers, together with an explicit rule defining when two frame-observer's accounts are mutually consistent or not.

\acknowledgements
Some ideas in this work were inspired by unpublished notes of R\"{u}diger Schack and Blake Stacey. I am grateful to Carlo Rovelli, Andrea Di Biagio and Federico Laudisa for answering many of my questions about the principles underlying RQM, although they probably disagree with the conclusions I have drawn from those principles. This work was supported in part by the John E. Fetzer Memorial Trust.

%\bibliographystyle{ieeetr}
%\bibliography{RQBisM.bib}

\end{document}